%% file: main.tex
\DocumentMetadata{}
\documentclass[sigconf]{acmart} % 2-column
\usepackage{makecell}
\usepackage{algorithm}
\usepackage{tikz}
\usepackage{algpseudocode}
\usepackage{amsmath}
\usepackage{algorithm,algpseudocode}

\definecolor{bcolor1}{rgb}     {1.0,0.0,0.0}
\definecolor{darkgreen}{rgb}     {0.0,0.5,0.0}
\definecolor{blue}{rgb}     {0,0.0,1.0}

\definecolor{red}{rgb}{1, 0, 0}
\definecolor{black}{rgb}{1, 1, 1}
\definecolor{blue}{rgb}{0, 0, 1}

\algnewcommand{\Inputs}[1]{%
  \State \textbf{Inputs:}
  \Statex \hspace*{\algorithmicindent}\parbox[t]{.8\linewidth}{\raggedright #1}
}
\algnewcommand{\Outputs}[1]{%
  \State \textbf{Outputs:}
  \Statex \hspace*{\algorithmicindent}\parbox[t]{.8\linewidth}{\raggedright #1}
}
\algnewcommand{\Initialize}[1]{%
  \State \textbf{Initialize:}
  \Statex \hspace*{\algorithmicindent}\parbox[t]{.8\linewidth}{\raggedright #1}
}

\AtBeginDocument{%
  \providecommand\BibTeX{{%
    \normalfont B\kern-0.5em{\scshape i\kern-0.25em b}\kern-0.8em\TeX}}}

\copyrightyear{2025}
\acmYear{2025}
\setcopyright{acmlicensed}\acmConference[IUI '25]{30th International Conference on Intelligent User Interfaces}{March 24--27, 2025}{Cagliari, Italy}
\acmBooktitle{30th International Conference on Intelligent User Interfaces (IUI '25), March 24--27, 2025, Cagliari, Italy}
\acmDOI{10.1145/3708359.3712114}
\acmISBN{979-8-4007-1306-4/25/03}

\begin{document}

%\title{Few-Shot Human-in-the-Loop Bayesian Optimization via Meta-Learning}
%\title[Few-Shot Human-in-the-Loop Optimization via Meta-BO]{Few-Shot Human-in-the-Loop Parametric Optimization \\for Wrist-based Pointing via Meta-BO}
\title[Redefining Affordance via Computational Rationality]{Redefining Affordance via Computational Rationality}
\author{Yi-Chi Liao}
\email{yichi.liao@inf.ethz.ch}
\orcid{0000-0002-2670-8328}
\affiliation{%
  \institution{ETH Zürich}
  \country{Switzerland}
}

\author{Christian Holz}
\email{christian.holz@inf.ethz.ch}
\orcid{0000-0001-9655-9519}
\affiliation{%
  \institution{ETH Zürich}
  \country{Switzerland}
}

\renewcommand{\shortauthors}{Liao and Holz}

\begin{abstract}
\input{tex/0.abstract.tex}

\end{abstract}

\begin{CCSXML}
<ccs2012>
<concept>
<concept_id>10003120.10003121.10003126</concept_id>
<concept_desc>Human-centered computing~HCI theory, concepts and models</concept_desc>
<concept_significance>500</concept_significance>
</concept>
<concept>
<concept_id>10003120.10003121</concept_id>
<concept_desc>Human-centered computing~Human computer interaction (HCI)</concept_desc>
<concept_significance>500</concept_significance>
</concept>
</ccs2012>
\end{CCSXML}

\ccsdesc[500]{Human-centered computing~HCI theory, concepts and models}
\ccsdesc[500]{Human-centered computing~Human computer interaction (HCI)}

\keywords{Affordance; Computational Rationality; Theory; Perception; Action; Reinforcement Learning; User Modeling; Design}



\maketitle

\input{tex/1.introduction.tex}

\input{tex/2.related_work.tex}

\input{tex/3.theory.tex}

\input{tex/4.thought_exp.tex}

\input{tex/6.new_foundation.tex}
\input{tex/7.discussion.tex}

\begin{acks}

The authors would like to thank Hee-Seung Moon, Antti Oulasvirta, Aini Putkonen, Christoph Gebhardt, Paul Streli, and Chieh-Ling Shih for their insightful discussions and feedback. We also thank the participants for their involvement in the thought experiments. This project was partly supported by the ETH Zürich Postdoctoral Fellowship Programme. 
\end{acks}

\bibliographystyle{ACM-Reference-Format}
\bibliography{sample-base}
\end{document}

%% file: tex/0.abstract.tex
Affordances, a foundational concept in human-computer interaction and design, have traditionally been explained by direct-perception theories, which assume that individuals perceive action possibilities directly from the environment. However, these theories fall short of explaining how affordances are perceived, learned, refined, or misperceived, and how users choose between multiple affordances in dynamic contexts. This paper introduces a novel affordance theory grounded in \emph{Computational Rationality}, positing that humans construct internal representations of the world based on bounded sensory inputs. Within these internal models, affordances are inferred through two core mechanisms: feature recognition and hypothetical motion trajectories. Our theory redefines affordance perception as a decision-making process, driven by two components: confidence (the perceived likelihood of successfully executing an action) and predicted utility (the expected value of the outcome). By balancing these factors, individuals make informed decisions about which actions to take. Our theory frames affordances perception as dynamic, continuously learned, and refined through reinforcement and feedback. We validate the theory via thought experiments and demonstrate its applicability across diverse types of affordances (e.g., physical, digital, social). Beyond clarifying and generalizing the understanding of affordances across contexts, our theory serves as a foundation for improving design communication and guiding the development of more adaptive and intuitive systems that evolve with user capabilities.

%% file: tex/1.introduction.tex
\section{Introduction}

Human beings have a remarkable ability to intuitively identify appropriate actions to take within their environments. 
For example, when seeing a mug, we instantly perceive that the mug affords us to grasp it, hold it, etc. 
Coined by Gibson \cite{gibson1966senses}, the term \textit{affordance} refers to ``the action possibilities that the environment presents.'' 
%While Gibson proposed a theory surrounding the term, he did not provide the exact perception mechanism of affordance. 
Rooted in his \textit{direct-perception} view, Gibson believes humans can directly ``pick up'' affordance information from the external world. 
Consequently, affordances and their corresponding perception are aligned. 
Within the framework, the definition and the perception mechanisms of affordance are unclear, or even missing, from the first place. 
%He never considered the case where affordance and its perception are mismatched.
%Because of these natures, the perception of affordance is ill-defined, barely explained, and poorly understood in Gibson's theory, which left future researchers to fill in. 
%Many debates surround this issue. 
%The most important question is ``what is affordance.''
To provide a concrete definition of affordance, ecological psychologists have developed divergent viewpoints. 
Property-based affordance characterizes affordance as an inherent property of the environment \cite{turvey1992affordances}, while relation-based affordance posits that affordance emerges from the relation between an actor and their environment \cite{chemero2003outline}. 
These views are conflicting \cite{chemero2003outline} and either provide a clear mechanism for perceiving affordances.

%These two schools of thought develop their definitions and the mechanisms of perceptions. 

Later on, the concept of affordance was introduced to the fields of design and human-computer interaction (HCI) \cite{norman1, 10.1145/108844.108856}. 
Due to the lack of concrete definition and perceiving mechanisms, researchers soon notice several significant challenges that the original affordance theory can not explain.
First, there are occasions when the perceived affordance deviates from the actual action possibilities, e.g., "false affordance" and "hidden affordance" \cite{10.1145/108844.108856}. 
The direct-perception theory can not explain this discrepancy between the physical world and perception. 
Norman attempted to answer this discrepancy by separating affordance and perceived affordance \cite{norman1988psychology}, but did not answer how we perceive affordance and why there are discrepancies. 
Second, there exist complicated actions that can not be easily explained by human's direct perception; for example, the ``save icon'' on an interface allows ``saving files'' \cite{mcgrenere2000affordances, norman1, 10.1145/108844.108856}. 
To address this issue, prior HCI researchers coined different terms, such as \textit{Technology or Digital Affordance} (action possibilities within the digital world) \cite{mcgrenere2000affordances}, \textit{Sequential Affordance} (a series of meaningful actions) \cite{10.1145/108844.108856}, \textit{Complex Affordance} (requiring previous experience or practice) \cite{mcgrenere2000affordances, albrechtsen2001affordances}, \textit{Social Affordance} (appropriate actions in certain social dynamics) \cite{10.1145/1274892.1274907}, etc.
While these terms expand the meaning of affordance, these deviations imply that there are multiple affordance definitions and perceiving channels, which are still unexplored or undefined \cite{mcgrenere2000affordances, gaver1996situating}.
Further, it shows that humans learn and adapt to new affordances, but the learning mechanism is unclear \cite{franchak2010learning}. 
Lastly, traditional affordance theories tend to detach affordances from the actor's internal goals and the decision-making processes that guide action selection, such as evaluating utility and confidence. 
Thus, they do not explicitly explain how humans make choices when faced with multiple affordances, especially in complex or ambiguous environments where goals and expected outcomes play a crucial role \cite{problem, bach2014affordance}.
In a nutshell, the affordance theories extended from Gibson's view can not provide concrete definitions of affordance or the perceiving mechanisms, leading to more ununified terms and confusion.
Further, these confusions hinder its actual application in design practice. 

\begin{figure*}
  \includegraphics[width=\textwidth]{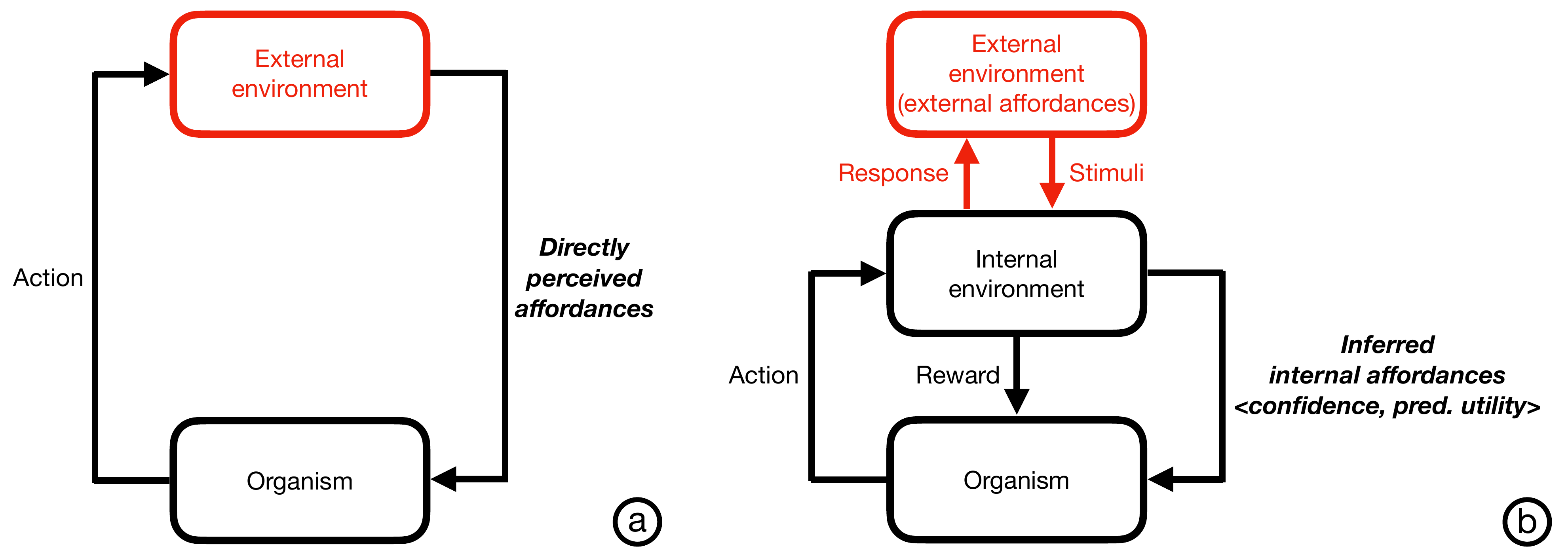}
  \caption{(a) Traditional affordance theory (Gibson's view): The organism directly perceives affordances as action possibilities presented by the environment, without the need for internal processing or cognitive mediation. These directly perceived affordances guide subsequent actions.
  (b) Computational Rationality-based affordance: Our theory proposes that while ``external affordances'' exist in the external environment, organisms can not directly perceive it. Instead, an internal representation (the ``internal environment'') is constructed from constrained sensory inputs. Within this internal environment, ``internal affordances'' are inferred, which contain two components: confidence in executing the action and predicted utility (expected outcome of taking an action). The organism refines these internal affordances through reinforcement learning, using feedback from previous actions (e.g., rewards) to continuously update and improve affordance inferences.}
  \label{fig:teaser}
\end{figure*}

We aim to fill in the gap.
Instead of inheriting Gibson's view in which affordance can be directly perceived as ``information pickup'', we argue that affordance perception is constructed, and can be computationally modeled. 
We present a novel affordance theory based on Computational Rationality (CR) \cite{gershman2015computational, oulasvirta2022computational} --- a framework that models human decision-making as an optimal process of reasoning and problem-solving under constraints in an internally constructed environment.
In CR, humans make decisions by evaluating actions based on their predicted outcomes, while accounting for the limited information and resources available. 
This framework emphasizes that decisions are not made based on direct access to external reality but through internal representations shaped by experience and sensory data.
In our CR-based affordance (illustrated in \autoref{fig:teaser}), we argue that humans (i.e., organisms) can not directly perceive information from the external world. 
Instead, we construct an internal representation of the world based on constrained sensory inputs. 
This mental construct, denoted as the "internal environment," is what humans have access to.
Because of the separation of external and internal environment, our theory deliberately segments between \textit{external affordance} and \textit{internal affordance}. 
External affordances refer to actions physically feasible for a given actor within an environment, which humans can not directly observe.
Internal affordances are humans' inferences within the internal environment. 
For each action, the internal affordance contains two pieces of information: the \textit{inferred confidence} of executing it and a corresponding \textit{predicted utility} of such an action. 
As an inference, the internal affordance may diverge from its external counterpart, i.e., this is when false or hidden affordances happen.
The predicted utility is shaped by the internal goal of the actor, and it strongly affects the final choice of action.
Take social affordance \cite{rietveld2013social} as an example, one can physically tap another person's head, but it is easy to predict this action comes with a negative utility, so rational humans will avoid taking this action despite the high confidence of successfully executing it. 

In our theory, two mechanisms are leveraged to infer internal affordances within the internal environment.
The first mechanism is \emph{feature recognition and matching}, where the features of the internal environment are compared against past experiences. 
For instance, the visual features of a mug align with other objects that afford you to ``grasp'' in the past, so you are able to perceive ``grasp'' affordance based on recognition.
The second mechanism involves \emph{generating hypothetical motion trajectories}. 
For instance, if confronted with uncertainty about passing through a narrow door, the mind can simulate the motion of traversing it. 
Through these mechanisms, individuals infer affordances with two key components: the confidence (how likely it will succeed) and utility (the predicted worth value of taking this action) attached to each action.
Our theory further explains the learning of affordance. 
When encountering new environments, one learns the right action through reinforcement learning --- simulating actions, predicting the action utility, executing actions, observing the actual utility, and iteratively improving.
The accumulated experience of affordance (action) and visual features allows humans to gradually rely on the feature recognition mechanism, enhancing recognition efficiency.
We argue that even the affordances that seem to be ``directly'' perceived (e.g., grasp a ball), were once learned through iterative trials. 
With this, our theory can effectively explain how humans learn, develop, and perceive the various types of affordances (e.g., digital and cultural affordances) in different fields. 
While these action possibilities are very different from the ones in the natural world, we still learn through reinforcement learning, and we recognize them through the internal representation of the world. 
In summary, this paper makes these key contributions:

\begin{itemize}
    \item \textbf{Novel CR-Based Affordance Theory}: We introduce a new \textit{CR}-based theory of affordance, shifting from the direct perception view to a framework involving \textit{internal representation} and \textit{decision-making}, including weighing \textit{confidence} and \textit{predicted utility}.
    
    \item \textbf{Unified Framework for Diverse Affordances}: Our theory serves as a unifying framework for various types and theories of affordances, including those developed in ecological psychology (e.g., property- and relation-based views) and those introduced in HCI and design (e.g., digital, social, and complex affordances). This clarified, unified perspective provides practical guidance for \textit{HCI}, \textit{user experience}, and \textit{interface design}.

\end{itemize}

%% file: tex/2.related_work.tex
\section{Related Work}

%We review the key works on affordance in ecological psychology, its relevance in design and HCI, and computational rationality. 

\subsection{Affordance in Ecological Psychology}

J. J. Gibson coined the concept of affordance as \emph{“the affordances of the environment are what it offers the animal, what it provides or furnishes, either for good or ill”} \cite{gibson1966senses}. Later, in \emph{The Ecological Approach to Visual Perception} \cite{gibson2014ecological}, he provided further detail: \emph{“If a terrestrial surface is nearly horizontal (instead of slanted), nearly flat (instead of convex or concave), and sufficiently extended (relative to the size of the animal) and if its substance is rigid (relative to the weight of the animal), then the surface affords support.”} To Gibson, affordances are action possibilities offered by the environment to animals. However, this concept remained broad and somewhat ambiguous, as Gibson did not clearly differentiate between affordances objectively present in the environment and those subjectively perceived by animals. This ambiguity stems from his theory of \emph{direct perception}, which posits that affordances are directly “picked up” from the environment without requiring cognitive mediation or internal representation.

Following Gibson, researchers in ecological psychology sought to refine the concept of affordance, leading to two distinct viewpoints. Some researchers view affordances as \emph{dispositional properties} of objects, inherent in the object's characteristics \cite{turvey1992affordances, reed1996encountering, michaels2000information}. These properties give rise to action possibilities always present in the environment. For instance, a chair's flat, stable surface affords sitting, whether or not it is currently in use. In this view, the environment “offers” these possibilities, which are actualized when an appropriate actor engages with the object.
Conversely, \emph{relation-based theories} argue that affordances arise from the \emph{interaction} between the actor and the environment \cite{heft2001ecological, chemero2003outline}. In this perspective, affordances are not static properties of objects but are dependent on the actor's capabilities and context. For example, a chair may afford sitting for an adult but not for an infant, as the action possibilities are shaped by the actor's size, strength, and situational context.

The two views present a fundamental conflict: \emph{property-based theories} view affordances as static, inherent possibilities in the environment, while \emph{relation-based theories} see affordances as dynamic and dependent on the actor's relation with the environment \cite{chemero2003outline}. 
%The challenge lies in reconciling whether affordances can be both objective properties and relational constructs.
Our \emph{CR-based theory} addresses this conflict by integrating both views. We propose that \textit{external affordances} exist as potential actions within the environment but are realized through the actor's internal processing. The perception of \textit{internal affordances} involves human goals, capabilities, and decision-making strategies, offering a unified approach to affordance perception. 

Furthermore, studies have shown that affordance perception improves with practice \cite{franchak2010learning, cole2013perceiving, day2017calibration}, but the learning mechanism has remained largely unexplored \cite{gibson2000perceptual, gibson2000ecological}. Our CR-based theory fills this gap by providing a concrete mechanism for affordance perception and learning through internal representations and decision-making. Actors infer affordances via \emph{feature recognition} and the \emph{simulation of hypothetical actions}, refined over time through \emph{reinforcement learning} \cite{sutton2018reinforcement, chater2009rational}. By integrating goals, confidence, and predicted utility, our theory explains affordance perception, learning, and action selection when faced with multiple possibilities.

\subsection{Affordance in HCI and Design}

Affordance is a foundational concept in HCI and design, with early pioneers like Norman \cite{norman1988psychology} and Gaver \cite{gaver1991technology}. However, confusion around the term arose almost immediately. Designers quickly encountered issues like ``hidden affordances'' (not perceptible) and ``false affordances'' (wrongly perceived) \cite{gaver1991technology}, which Gibson's direct perception theory could not fully explain. Norman addressed this by distinguishing between affordances (how objects can be used) and perceived affordances (how users interpret these uses). However, this separation introduced complexity, implying that perception could misalign with actual object properties. Norman's focus on perception moved the discussion away from Gibson's idea of inherent affordances, leading to fragmented views across HCI \cite{mcgrenere2000affordances}.

As technology advanced, new affordances emerged in digital environments. Researchers introduced terms like \emph{digital/technological affordance} \cite{gaver1991technology}, describing how interface design shapes user interactions, and \emph{sequential affordances}, which unfold over multiple steps, such as completing a task. 
Turner distinguished between \emph{simple affordances} (aligned with Gibson's original ideas) and \emph{complex affordances}, which rely on learned interactions \cite{turner2005affordance}. 
Expanding further, \citet{ramstead2016cultural} introduced \emph{cultural affordances}, based on social norms. More recent research has explored \emph{activity-based affordances}, which emphasize the social-historical context of interactions \cite{albrechtsen2001affordances,baerentsen2002activity,kaptelinin2014affordances}. These concepts highlight how action possibilities are shaped by more than just physical properties, considering culture, history, and context.

Given the variety of definitions, there is a need for a unified theory to reconcile these interpretations in HCI and design. As \citet{mcgrenere2000affordances} noted, HCI researchers have followed different streams of affordances: either Gibson's original concept \cite{ackerman1996zephyr,bers1998interactive,zhai1996influence,vicente1992ecological}, Norman's idea \cite{conn1995time,johnson1995comparison,nielsen1997user}, or new variations \cite{mohageg1996user,shafrir1994visual,vaughan1997understanding, 10.1145/3491102.3501992}. These differences suggest that affordances are perceived and learned through various approaches, making unification crucial for a clearer understanding and better design applications.
Our theory fills in this gap by the concept of Computational Rationality. A deeper analysis of this is presented in \autoref{Sec:Variations}. 
%Building on Norman's distinction between affordances and perceived affordances, we utilize Computational Rationality to introduce a more detailed cognitive process of affordances learning. Rather than treating affordances as static or separate from perception, our theory focuses on how users dynamically perceive and adapt to affordances through mechanisms like feature recognition and reinforcement learning. By considering both external object properties and internal user goals and experiences, our theory offers a comprehensive framework for affordance perception in HCI.

\subsection{Computational Rationality}

Computational Rationality (CR) is a framework for modeling decision-making as an optimal process constrained by the cognitive and environmental limits individuals face. Unlike classical rationality, which assumes perfect information and unlimited processing capacity, CR accounts for the bounded nature of human cognition, including limited time, memory, and cognitive effort. Built on Herbert Simon's foundational concept of \emph{bounded rationality} \cite{simon1955behavioral}, it emphasizes that human decision-making is shaped by the need to make satisfactory, rather than perfect, decisions within these cognitive constraints.
CR extends Simon's ideas by viewing decision-making as constrained by both environmental factors and cognitive limitations. Human decisions, while not optimal in a classical sense, are considered near-optimal when accounting for these constraints \cite{gershman2015computational}. The process balances trade-offs between speed, accuracy, and cognitive effort, relying on experience, heuristics, and mental models formed through learning \cite{griffiths2015rational}.

CR revolves around internal representations and predictive models, which combine sensory input with learning. These models help generate expected utility, a principle in decision theory, to predict action outcomes based on past experiences \cite{simon1955behavioral, russell2016artificial, savage1972foundations}. In CR, decision-making involves selecting actions that maximize utility within resource constraints like time, cognitive effort, and sensory data \cite{griffiths2015rational, kahneman2003maps}. Reinforcement learning enables individuals to refine these models by iterating based on feedback from the environment \cite{sutton2018reinforcement, chater2009rational}. Over time, successful actions are reinforced, shaping future decisions and allowing users to adapt to changing contexts \cite{dayan2008decision, daw2005uncertainty}. This process supports the idea that learning is central to perceiving affordances, or action possibilities \cite{gershman2015computational, daw2005uncertainty, shadmehr2008computational}.

Oulasvirta et al.'s work extends these concepts into HCI \cite{oulasvirta2022computational}, proposing that human-computer interactions are shaped by decision-making processes governed by bounded rationality.
However, this paper does not explicitly address affordances. 
%Instead, it explores how users adapt behavior to interfaces and how designs can optimize for human limitations.
Our theory fills in the gap by using CR as the basis to explain affordance perception and learning. 
We propose that affordances are not directly perceived but are inferred through cognitive processes shaped by constraints. 
By incorporating expected utility and reinforcement learning --- key components of CR, our theory explains how users infer affordances, and refine them through feedback. 

\subsection{Affordance in Machine Learning}
Computer vision and robotics fields have explored various types of computational affordance models, aiming to enable robots to efficiently select actions in a complex environment.
Affordance recognition is primarily seen as a classification problem where deep neural nets are employed to train on labeled images \cite{nguyen2017object, AffordanceNet18, chuang2018learning, van2016stable, finn2016deep, 7051290,7139369,KJELLSTROM201181, 10.1007/978-3-540-79547-6_42}. 

More recent works have integrated reinforcement learning (RL) and motion planning, allowing robots or agents to explore and learn affordances through interaction with their environment. Notable works include \citet{nagarajan2020learning}, who proposed an RL-based approach where robots discover the appropriate pre-defined actions for interacting with objects through trial and error. \citet{5995327} further explored affordance computation by generating motion trajectories to interact with objects, while \citet{10.1145/3491102.3501992} introduced a reinforcement learning-based affordance model where agents learn the appropriate motions from scratch via RL.
Our theory builds on these works, which view affordance as something learned and recognized through motion planning. However, we take this further by framing affordance inference as a dynamic, decision-theoretic cognitive construct. 
Unlike previous RL models, which focus on \emph{affordance learning and detection}, we aim to contribute a more holistic and unifying \emph{theory} and \emph{definition}. 
Specifically, our theory distinguishes between \textit{external affordances} and \textit{internal affordances} and introduces the concepts of \textit{confidence} and \textit{predicted utility} as core elements of affordance perception, which previous models do not consider at all. Our generalized framework further provides actionable insights for HCI and design practice.

%% file: tex/3.theory.tex
\section{CR-based Affordance: Theoretical Commitment}

We propose a new affordance theory grounded in Computational Rationality (CR), offering a concrete framework for affordance perception and learning. Unlike traditional theories like Gibson's, which emphasize direct perception, we argue that affordances are constructed within an internal representation of the world. This internal model is shaped by limited sensory inputs, cognitive constraints, and individual experiences. 
Therefore, in our theory, affordances is understood as two parts: external affordances, representing action possibilities in the environment, and internal affordances, which reflect the individual's inference of those possibilities.

We further propose that each internal affordance contains two components: Confidence (the perceived likelihood that the action can be successfully executed) and Predicted Utility (the expected value or outcome of the action in relation to the individual's goals and priorities).
These internal affordances are continuously updated through mechanisms of feature recognition and hypothetical motion trajectories, which allow organisms to evaluate potential actions in both familiar and novel situations. 
By integrating reinforcement learning into this framework, our theory explains how individuals improve their ability to perceive and act on affordances over time. Even affordances that seem to be directly perceived are learned and refined through repeated interaction with the environment. This approach offers a unified, computational framework that can account for a wide range of affordances, from simple physical actions to more complex, context-dependent possibilities.

With this framework, we introduce our theoretical commitments,
each of them is further elaborated on in the following subsections: 

\begin{enumerate}
    \item \textbf{Bounded Optimality:} Organisms (or agents) solve bounded optimality problems by selecting the best action possibility inferred in the internal environment based on what is offered by the external environment.
    \item \textbf{External Affordances:} These are objective, physical action possibilities in the external environment that are not directly accessible by the organism.
    \item \textbf{Internal Affordances:} Internal affordances refer to action possibilities inferred within an internal representation of the external environment that organisms construct.
    \item \textbf{Affordance Inference:} Internal affordance inference consists of two elements: confidence (the likelihood of successfully performing the action) and the expected utility of that action.
    \item \textbf{Mechanisms of Inference:} Internal affordances are inferred through two processes -- (a) feature recognition and matching and (b) hypothetical motion trajectories.
    \item \textbf{Affordance Learning:} Internal affordances are continuously learned and refined through reinforcement, shaped by interaction experience with the external environment.
\end{enumerate}

\subsection{Organism (or agent)}

Following ecological psychology, we use the term organism to refer to any actor situated within an environment. 
In the context of CR, this entity is commonly referred to as an agent. These two terms (organism and actor) are interchangeable in later content. 
The organism constructs its internal environment based on limited and constrained sensory inputs (i.e., stimuli) from the external world and infers multiple internal affordances within this cognitive space. 
The organism's goal is to select the action (i.e., response) with the highest predicted utility and highest confidence, thus adhering to the principles of bounded optimality (Commitment 1).

\subsection{External Environment and External Affordance}

The \textit{external environment} represents the objective, physical world that exists independently of any individual's perception. \textit{External affordances} refer to the action possibilities physically present in this environment. 
These affordances are inherent properties of objects or environments and remain consistent regardless of whether or not they are perceived by an actor (Commitment 2). 
For example, the flat, stable surface of a chair affords sitting, irrespective of whether an individual recognizes or chooses to sit on it.
From the view of CR, external affordances are not directly observable by the organism. The organism instead perceives and acts upon internal affordances, which are subjective inferences about these external possibilities.

\subsection{Internal Environment and Internal Affordance}

In contrast, the \textit{internal environment} is a cognitive representation of the external world, constructed based on constrained sensory input and shaped by past experiences, memory, and learning. 
It can be constructed in different forms. For instance, it could be a visualizable model of the environment. For example, when visually seeing a room and forming a mental image of it.
It could also be an abstract, rule-based model where prior experiences dictate expectations and allow for predictions. For instance, upon hearing the sound of a plate falling, one can infer it might break without needing to mentally visualize the scene.
Within this internal environment, \textit{internal affordances} represent the subjective inferences about action possibilities (Commitment 3). 
These affordances emerge through cognitive processes and could diverge from their external counterparts, leading to phenomena like false affordances (when an action appears possible but the action cannot actually be executed) and hidden affordances (when an action is possible but is not identified by the organism due to lack of cues).

\subsection{Affordance Inference (Confidence and Predicted Utility)}

Traditional affordance theories often see affordances as binary; that is, an action can or cannot be performed \cite{gibson1977theory, gibson1966senses} or confidence level; i.e., how likely it is that the action can be successfully executed \cite{franchak2014affordances}. 
While useful in simple contexts, such as physical affordance, these approaches do not account for the complexity of human decision-making in real-world environments, such as social affordance or more complex decision-making. 
To enable the most generalizability, we extend the model to include two critical components (Commitment 4):

\begin{itemize}
    \item \textbf{Confidence}: The actor's belief in how likely they are to successfully execute the action.
    \item \textbf{Predicted Utility}: The expected value or outcome of attempting the action based on the actor's internal goals and priorities.
\end{itemize}

Thus, we represent internal affordances as a tuple: \textbf{<action, confidence, predicted utility>}. Both confidence and predicted utility are dynamic, continuously updated as the actor receives new sensory inputs or learns from experience. These two elements are essential in shaping decision-making: even if an actor is highly confident that they can perform an action, they may still choose not to act if the predicted utility (i.e., the potential outcome) is negative. 
Similarly, if the predicted utility is high but confidence in success is low (e.g., in high-risk, high-reward situations like gambling), the actor might hesitate to act.

This model helps to clarify that affordance perception is not simply about estimating physical capability but also involves complex decision-making involving a continuous spectrum between ``able or unable'' (e.g., physical affordance in simple scenarios) and ``willing or unwilling'' (e.g., social affordance). 
Previous theories often blurred this distinction, complicating attempts to explain the coexistence of different affordance types. 
Our framework explicitly separates physical feasibility (confidence) from motivational factors (predicted utility), offering a better understanding of how humans evaluate action possibilities.

This distinction is particularly important for explaining more complex forms of affordances, such as social, cultural, and collective affordances. 
In collective affordances \cite{weichold2020collective}, for instance, confidence in initiating a cooperative action (e.g., jointly lifting a heavy object) is influenced not only by one's physical ability but also by expectations of others' abilities. 
In cultural affordances \cite{ramstead2016cultural}, actions are shaped by norms -- what might be physically possible could be socially inappropriate, meaning predicted utility is key in decision-making. 
For example, offering a handshake may afford a positive interaction in one culture but could be viewed as inappropriate in another, shifting the utility of that action.

By incorporating both confidence and predicted utility into the affordance model, we can explain not only whether an action can be performed, but also why individuals choose or avoid certain actions based on their goals, experiences, and social contexts. This theory provides a more generalized and comprehensive understanding of affordance perception and action selection across a variety of real-world situations.

\subsection{Mechanisms for Inferring Internal Affordances}

Extending the existing computational affordance models, which detect affordance via feature recognition \cite{nguyen2017object, AffordanceNet18, chuang2018learning} or via motion planning \cite{nagarajan2020learning, 10.1145/3491102.3501992}, our CR-based theory argues that organisms rely on both mechanisms to infer the internal affordances in the internal environment (Commitment 5):

\begin{itemize}
    \item \textbf{Feature Recognition and Matching}: Organisms match features of the internal environment to prior experiences, allowing them to recognize affordances based on similarity. For example, recognizing that a mug affords grasping due to its similarity to other graspable objects.
    \item \textbf{Hypothetical Motion Trajectories}: In cases of uncertainty, organisms simulate possible actions within their internal environment, mentally projecting how an action might unfold. This allows them to estimate both the confidence and utility of an action before attempting it, enhancing the result of decision-making.
\end{itemize}

In cases where the internal environment is visualized, \textbf{feature recognition and matching} becomes a straightforward process. When an actor visually perceives an object or imagines a specific mug, their internal environment generates a mental representation that mirrors the object's visual characteristics, allowing for more immediate recognition of affordances. 
For example, when seeing a mug, the internal environment mirrors its shape, handle, and size. The actor quickly matches these visual features to previously encountered objects, allowing them to recognize the mug's affordance for grasping. 
Visual feature matching allows for direct and efficient affordance inference, especially in contexts where the physical properties of objects are visually apparent and easily linked to past experience.

Different from \citet{10.1145/3491102.3501992}, we argue that an internal environment does not need to be visualized but can also be an abstract representation of the external world, where rules and past experiences guide the recognition of action possibilities. 
Even in this abstract form, \textbf{feature matching} remains a critical mechanism. 
For example, hearing the sound of a plate falling in another room, without visualizing the room, allows an individual to infer affordances such as the likelihood of the plate breaking and the danger of sharp fragments on the floor, based on the feature match to past experiences (e.g., ``falling plates often break'', and ``broken plates result in sharp fragments''). 
Similarly, when hearing the word "doorknob," one can instantly infer its affordances, such as grasping or opening a door, by matching the term to previously learned features and functionalities associated with this term, which does not need to visualize a specific doorknob. 
This allows the actor to infer affordances even in situations where direct sensory information is limited or unavailable.

On the other hand, \textbf{hypothetical motion trajectories} require the internal environment to be visualized and used to simulate possible motion trajectories. This mechanism becomes especially important when actors encounter unfamiliar or uncertain scenarios where the affordances are not immediately recognizable through feature matching alone. For example, if an individual is unsure whether they can pass through a narrow doorway while carrying a wide object, they might mentally simulate the movement --- projecting their body's motion and the object's size relative to the doorway. Through this mental simulation, they can estimate the confidence (how likely the action will succeed) and the predicted utility (whether passing through the door is useful or desirable).

These mechanisms often work together. In familiar environments, feature recognition alone may suffice for affordance perception. Yet, in more uncertain or novel situations where feature recognition is inadequate, hypothetical motion trajectories help actors explore the outcomes of possible actions before they are taken.
%, thus updating confidence and predicted utility.

\subsection{Affordances are Learned and Refined Through Reinforcement}

Affordance perception is not innate but learned through an iterative \textit{reinforcement learning} process (Commitment 6). When organisms receive sensory input from the external world, they construct an internal environment, infer possible affordances, select the optimal action, and observe the outcome. This outcome includes the \textit{result} (whether the action succeeded or failed) and the \textit{actual utility}. The actual utility serves as a \textit{reward signal} in reinforcement learning, guiding future decisions and affordance perceptions.

Three key elements -- (1) the internal environment, (2) confidence, and (3) predicted utility -- are continuously shaped and improved based on these reward signals. Positive outcomes reinforce current inferences, while negative outcomes drive updates to future predictions. Below, we explain each element:

First, the internal environment, which is a mental model of the external world, is updated through feedback. For instance, when encountering a box on the floor, the internal environment might initially simulate a successful lift. Upon attempting and failing due to the box's weight, the internal environment updates to reflect the new information. In future encounters, the individual's internal model might simulate failure or suggest alternative strategies, such as seeking assistance or trying a different approach.

Second, confidence is recalibrated through experience. If an individual approaches a door assuming it is fully openable but discovers it is locked or stuck, this experience adjusts their internal environment's future predictions. The next time they encounter a similar door, they may approach with lower confidence or anticipate additional factors like checking the lock first. This learned adjustment ensures that the internal model better matches real-world outcomes.

Third, predicted utility is continuously refined based on outcomes. If the goal is to open a door, successfully doing so leads to a positive reward signal, reinforcing the decision to act. 
Conversely, failing to open the door is an undesirable outcome, which is perceived as a negative reward, thereby discouraging the same action in the future. 
This feedback loop helps the individual fine-tune both confidence and utility estimates in the future.

Reinforcement learning applies not only to \textit{hypothetical motion trajectories} but also to \textit{feature recognition}. Repeated interactions with specific features -- such as doors that will not open -- can lead to the automatic recognition of affordances without the need for simulating actions. This process allows for more efficient decision-making and resembles what Gibson described as “direct perception,” where individuals can immediately perceive the affordances of their environment without deliberation. However, in contrast to Gibson's view of affordances as being directly perceived from the environment, our theory suggests that such immediate recognition is a learned ability. 

Together, this dynamic feedback loop continually refines the perception of affordances. Through repeated interactions, successful actions are reinforced, and unsuccessful attempts update inferences, refining internal models, confidence, and utility estimates. 
This iterative process ultimately enables more efficient affordance inference and better-informed decision-making over time.

%% file: tex/4.thought_exp.tex
\section{Thought Experiments}

We validate key elements of our CR-based affordance via a series of thought experiments.
8 participants (age 25 - 30, including 2 females) were recruited to conduct the experiment; they were instructed to answer the questions and provide rationales. 

\subsection{Thought Experiment 1: Internal Affordance}

This thought experiment challenges a fundamental aspect of Gibson's affordance theory --- that humans directly pick up affordance information from the external world. In contrast, our theory posits that we do not have direct access to the external environment, nor do we perceive affordances as direct features of that environment. Instead, we construct an \textit{internal environment}, a representation of the world based on bounded sensory input, and detect affordances within this internal environment.

\paragraph{Thought Experiment: Breakability of an egg in different contexts}

Imagine you are standing in a room with an egg in your hand. Do you think the egg is breakable? Your answer is most likely yes. Now, imagine the same scenario, but this time the entire room, including yourself, is fully covered in a thick sponge. Do you still think the egg is breakable?

\paragraph{Analysis}

In both scenarios, your perception of the egg's affordance (breakability) is based on the internal environment you have constructed in your mind. In the first scenario, without any external sensory input or visual stimuli, you detect the affordance of the egg's breakability within your internal representation of the environment. The egg, in this imagined context, is perceived as breakable.
In the second scenario, where the room and you are covered in thick sponges, the internal environment you construct has changed. As a result, your perception of the egg's breakability is now different. The thick sponge in the internal environment affects how you evaluate the affordance of the egg, leading you to question whether it can still break. Even though no external physical properties of the egg have changed, your internal representation of the environment shapes how you detect its affordances.

This experiment illustrates that affordances are inferred within the internal environment rather than directly perceived from the external world. Changes to that internal environment (e.g., the sponge) can shift how you assess the affordance, even if the object's physical properties remain constant.

\paragraph{Empirical Results}

In the first scenario, all eight participants responded that the egg was breakable. In the second scenario, six participants changed their answer (the egg was no longer breakable) while two maintained their original answer. When asked how they judged the egg's breakability, all participants mentioned they relied on general knowledge (e.g., that eggs are fragile) and past experiences (e.g., accidentally breaking eggs). 
Six participants specifically described constructing a visualized internal environment, imagining the room and egg in detail before making their decision. The other two participants, who did not explicitly mention visualizing the room, still arrived at their answer by applying past experiences and physical rules, indicating the use of a more abstract, rule-based internal environment. 
This supports our theory that internal affordances can be inferred through either a visualized internal environment or a more abstract, rule-based environment.
The results also show the key mechanisms of feature matching (e.g., eggs can break) and hypothetical motion trajectories (e.g., imagining the room is physical properties).
Overall, these findings align with our theory that affordances are inferred not directly from external stimuli but through the constructed internal environment.

\subsection{Thought Experiment 2: Confidence and Predicted Utility}

This thought experiment explores the roles of \textit{confidence} and \textit{predicted utility} in affordance inference, showing that these factors are critical in evaluating action possibilities. We present two sub-thought experiments to illustrate these concepts: one focused on confidence and the other on predicted utility.

\subsubsection{Thought Experiment 2.1: Change of Confidence}

You are offered a chance to buy a lottery ticket (\$1) that has a 1-in-1,000 chance of winning \$100. Do you think you will win? And will you buy the lottery?
Now, imagine you are offered a different lottery ticket with a 1-in-10 chance of winning the same \$100. Do you think you will win now? And will you buy the lottery now?

\paragraph{Analysis}

In the first scenario, the odds are stacked against you (1-in-1,000), leading to low confidence in winning the prize. The expected utility, calculated as \$100 multiplied by the probability of winning (1/1000), is only \$0.10. This low expected value, combined with low confidence, makes the action of buying the ticket unappealing.

In the second scenario, the odds improve significantly to 1-in-10, which increases your confidence in winning. The expected utility also rises to \$10 (1/10 $\times$ \$100), making the action more attractive. The increased confidence further enhances the expected utility, making it more likely that you will buy the ticket. This experiment illustrates how confidence and expected utility together influence decision-making and perception of affordance.

\paragraph{Empirical Results}

In the first scenario, all eight participants responded that they would not expect to win and would not buy the lottery ticket. In the second scenario, five participants still expected to lose, but three indicated uncertainty. Yet, all participants reported that they would buy the ticket in the second scenario, highlighting a shift in their decision-making process.

This shift in action highlights how both confidence and expected utility jointly impact the decision to act and perception of affordance. As the confidence improved from 1-in-1,000 to 1-in-10, participants' confidence in winning increased, and the expected utility rose from \$0.10 to \$10, making the decision to buy the ticket more appealing. As one participant (P3) noted, \emph{``The change in odds significantly boosts my belief in winning, and the expected return is higher than before.''}

\subsubsection{Thought Experiment 2.2: Change of Predicted Utility}

Imagine you are standing on a narrow plank suspended between two buildings, and there is \$100 on the other side. The plank is sturdy, and the distance between the buildings is only a few feet. Below you is a safety net, ensuring that if you fall, you will not get hurt. Do you think you can cross the plank?
Now, imagine the same scenario, but this time there is no safety net, and the drop below is several hundred feet. The plank remains sturdy, and the distance between the buildings is still only a few feet. Do you think you can cross the plank now?

\paragraph{Analysis}

In both scenarios, your confidence in walking across the plank remains largely the same because the physical conditions of the plank and the distance between the buildings have not changed. You likely believe you have the physical ability to walk across in both cases, meaning your confidence in the action (crossing the plank) is stable.
However, the predicted utility varies significantly between the two scenarios. In the first scenario, with the safety net, the risk is low, and the potential negative outcome (falling) has minimal consequences. This increases the predicted utility of walking across the plank, as the potential reward (\$100) outweighs the minor risk.
In the second scenario, without the safety net and with a dangerous drop below, the predicted utility decreases dramatically. Even though the plank is just as sturdy, the risk of severe injury or death outweighs the reward of \$100, making the action less desirable despite your confidence in your ability to cross.
This thought experiment shows how predicted utility --- the perceived value of the outcome --- affects decision-making. Even with high confidence, a negative predicted utility can deter you from taking action.

\paragraph{Empirical Results}
In the first scenario, all participants reported they could cross the plank. In the second scenario, five participants changed their response, saying they could not cross, while three still believed they could but chose not to.
This shift illustrates that predicted utility drives decision-making. 
While physical confidence in crossing the plank remains, the risk in the second scenario (falling without a safety net) lowered predicted utility, causing most participants to reconsider their actions.

\paragraph{Summary of this Experiment}
These two sub-thought experiments illustrate the interplay between confidence and predicted utility in affordance perception. In the first experiment, the change in confidence and utility affects your willingness to act. In the second experiment, despite stable confidence, the change in predicted utility (due to varying consequences) influences your decision to act. Together, these examples highlight how both factors are evaluated in the internal environment to shape action possibilities.

\subsection{Thought Experiment 3: Two Mechanisms for Inferring Internal Affordances}

This thought experiment demonstrates the two mechanisms by which internal affordances are inferred: feature recognition and hypothetical motion trajectories. 
According to our theory, when encountering an object or situation, individuals either rely on recognizing familiar features from past experiences or simulate possible actions to predict the outcome and assess the affordance.

\paragraph{Thought Experiment: Crossing a Stream via Rocks}

Imagine you are standing in front of a shallow but fast stream with several large rocks spaced across it. You need to cross the stream to get to the other side. The rocks look flat and stable and they are well-connected to each other. Do you think you can step on the flat rocks to cross the stream?
Now, imagine that the rocks are uneven, wet, and spaced further apart. You are not sure if you can reach the next rock without slipping or falling. Do you think you can cross the stream now?

\paragraph{Analysis}

In the first scenario, you can infer the affordance of stepping on the flat rocks by using feature recognition. Based on your past experiences with similar rocks, you recognize their flat surfaces as stable, affording the possibility of stepping on them safely to cross the stream. The action is clear because you have encountered similar situations before, and feature recognition provides enough information to evaluate the affordance without further mental simulation.

In the second scenario, the rocks are uneven and spaced farther apart. Here, feature recognition alone may not be sufficient to determine whether you can cross the stream. In this case, you would likely engage in hypothetical motion trajectories, mentally simulating the action of stepping onto the uneven rocks and predicting whether you can maintain balance or reach the next rock. You imagine the movement, assess your confidence in successfully completing the action, and evaluate the predicted utility of crossing the stream. This mental simulation helps you infer whether the affordance exists, even in the absence of immediate experience with this specific configuration of rocks.

\paragraph{Empirical Results}

All participants reported utilizing feature matching as part of their inference process, comparing key features of the rocks (flat, dry, well-connected) to past experiences. This allowed them to infer that crossing was possible in the first scenario. However, in the second scenario, where the rocks were uneven and spaced farther apart, seven participants noted that feature recognition alone was not enough, and they engaged in hypothetical motion trajectories. They described mentally simulating their movements, imagining the distance between rocks, and estimating the risk of slipping, allowing them to determine whether crossing was feasible. One participant relied primarily on feature recognition and matching, focusing on the perceived physical properties of the rocks.
These results support the two core inference mechanisms in our theory. In more familiar or certain situations, feature recognition suffices for affordance evaluation, while in uncertain contexts, hypothetical motion trajectories play a critical role in assessing potential actions. 

\subsection{Thought Experiment 4: Affordances Learned and Refined through Reinforcement}

This thought experiment demonstrates how affordances are learned and refined through reinforcement by interacting with an environment where feedback from actions helps shape future decisions. Our theory posits that individuals adjust their perception of affordances through trial and error, gradually refining their internal models based on the success or failure of their actions.

\paragraph{Thought Experiment: Navigating a Series of Doors}

Imagine you are inside a unique house with a series of doors, each different from the next. As you explore the house, you approach the first door. It is stiff, but with some effort, you manage to push it open. What does this teach you about the next door you encounter?

Now, you come to a second door that looks similar to the first. 
This door is even stiffer, and despite your repeated attempts, you cannot get it to budge. What will you do when you encounter the third door?

\paragraph{Analysis}
When you push open the first door, you learn that doors in this house may afford opening but require effort. This feedback adjusts your internal model, suggesting that similar doors might behave the same way.
However, the second door, despite looking similar, doesn't open with more force. This failure updates your internal model, showing that not all doors afford opening, even if they appear alike. The third door now represents a scenario where your actions are guided by past experiences --- knowing some doors might not be openable.

This illustrates how affordances are refined through reinforcement learning. As you interact with each door and receive feedback, your internal model of the environment evolves, allowing you to better predict the confidence in executing certain actions and the potential utility.

\paragraph{Empirical Results}

All eight participants reported expecting similar levels of effort for the second door and assumed it would be openable. However, when the second door was found to be non-openable, three participants adjusted their expectations for the third door, stating they would not consider it openable. Three participants reported being unsure, while only two participants still believed the third door would be openable.

These observations align with our theory that affordances are refined through reinforcement learning. As participants received feedback on their actions, their internal models evolved, allowing them to better assess the likelihood of success (confidence) and the utility of attempting further actions.

\subsection{Overall Findings}

Throughout four thought experiments, we demonstrated empirical support for our CR-based affordance theory with these insights: (1) affordances are inferred within a constructed internal environment rather than directly perceived from external stimuli, (2) the perception of affordances and the decision to act are jointly influenced by confidence and predicted utility, (3) internal affordances are inferred through two core mechanisms -- feature matching and hypothetical motion trajectories, and (4) affordances are refined and updated through feedback and reinforcement learning. 
Together, these findings validate our theory, showing that affordance perception is a dynamic and evolving process shaped by both internal cognitive models and external feedback.

%% file: tex/6.new_foundation.tex
\section{CR-based Theory as a Unifying Framework for other Affordance Variations}
\label{Sec:Variations}
Our \textit{CR-based affordance theory} provides a unifying framework for understanding a wide range of affordances by incorporating \textit{learning}, \textit{internal representation}, and \textit{decision-making}. 
%This section highlights several key affordance types and explains how our theory accounts for the process through which users perceive, learn, or refine their understanding of these affordances.

\subsection{Affordance in Ecological Psychology}

We first review how our theory extends and completes the affordance theories developed in \textit{Ecological Psychology}. 
After Gibson coined the term affordance, researchers in Ecological Psychology have developed two distinct perspectives: property-based and relation-based theories. 
%emphasize that affordances emerge from the interaction between an individual’s capabilities and the environment. 
%These views have often been seen separately and even seen as conflicting. 
We will specifically show how our CR-based theory fills this gap by integrating both perspectives, providing a more comprehensive model of affordance perception.
%We will also explore how our theory complements broader ecological affordances.

\vspace{5px}
\noindent
\textbf{1. Gibson's Original Affordance}\\
\textbf{\textit{Definition:}} Affordances are action possibilities based on the physical properties of objects \cite{gibson1966senses}. \\
\textbf{\textit{Example:}} A door handle affords pulling. \\
\textbf{\textit{Our Theory:}} Users initially construct an internal model of the environment based on information through sensory channels. Then they detect physical affordances through \textit{feature recognition} within the internal environment, relying on prior experiences to recognize similar objects and actions. If the affordance is unfamiliar or uncertain, they can use \textit{hypothetical motion trajectories} to simulate the action, predicting both the confidence of success and the expected utility of the outcome. As users interact with physical objects, their internal models are refined through reinforcement learning, improving the accuracy of affordance detection over time.

\vspace{5px}
\noindent
\textbf{2. Property-Based Affordance}\\
\textbf{\textit{Definition:}} Affordances are inherent properties of an object, such as its shape, texture, or material. These affordances exist independently of the observer and are intrinsic qualities of the object \cite{turvey1992affordances}. \\
\textbf{\textit{Example:}} A chair affords sitting because of its physical properties --- flat surface, height, and stability. \\
\ \textbf{\textit{Our Theory:}} In our CR-based framework, property-based affordances are understood as \emph{external affordances} present in the physical world. Organisms perceive these properties through sensory channels and then detect the corresponding \emph{internal affordances} within their constructed internal environments. For example, the flat, stable surface of a chair physically affords sitting (external affordance), while users perceive this property visually and construct an internal model where the sitting affordance is inferred.

\vspace{5px}
\noindent
\textbf{3. Relation-Based Affordance}\\
\textbf{\textit{Definition:}} Affordances are emerged from the relationship between the actor and the environment. Thus, affordances are \emph{not} fixed properties of objects but determined based on the actor’s capabilities relative to the object \cite{chemero2003outline}.\\
\textbf{\textit{Example:}} A chair might afford sitting for an adult but not for an infant, or a tall shelf affords reaching only for someone with the height or a tool to extend their reach.\\
 \textbf{\textit{Our Theory:}} In our CR-based framework, relation-based affordances correspond to \emph{internal affordances}. Our theory proposes that within the internal environment, the organism infers affordances based on its capabilities and goals, evaluating the confidence and utility of taking an action. For example, when considering a tall shelf, the internal model helps the user simulate actions (e.g., standing on a stool to reach an item) and assess the confidence in successfully completing the action. 
This dynamic interaction between the user’s capabilities and the environment’s affordances reflects the relation-based nature of affordance perception.

\vspace{5px}
\noindent
\textbf{4. Ecological Affordance}\\
\textbf{\textit{Definition:}} Ecological affordances refer to the action possibilities that emerge not from isolated objects, but from the interaction between an organism and a larger, integrated ecological system. These affordances account for how multiple elements in the environment collectively create opportunities for action \cite{chemero2013radical}.\\ 
\textbf{\textit{Example:}} A public park affords multiple actions like walking, sitting, and socializing, with trees providing shade, paths allowing movement, and benches offering rest. They work together to create a cohesive affordance system.\\
\textbf{\textit{Our Theory:}} In the CR-based theory, ecological affordances are perceived by constructing an internal model that considers the environmental system as a whole, rather than focusing on individual objects. Organisms infer how different elements within the system (e.g., trees, benches, and pathways) interact to afford actions. Using feature recognition, organisms detect key environmental elements, and through hypothetical motion trajectories, they simulate complex interactions across different objects. As they interact with different components, organisms receive feedback that helps refine their internal models, allowing them to perceive affordances for the environment as a collective.

\subsection{Affordance in HCI and Design}

Since affordance was introduced to HCI and design, they have expanded in meaning and terminology \cite{mcgrenere2000affordances}. We show how our CR-based theory provides a unifying and generalizable framework for this evolving spectrum of affordances.

\vspace{5px}
\noindent
\textbf{5. Perceived Affordance}\\
\textbf{\textit{Definition:}} Perceived affordances are what users believe are present based on their perception of an object or interface, which may diverge from the actual affordance \cite{norman2013design}. \\
%While Norman did not explicitly separate perceived affordances from actual affordances, his discussion of design cues like buttons and handles implies this distinction \cite{problem}. \\
\textbf{\textit{Example:}} A button on a screen looks like it can be clicked, even if it’s not actually functional. \\
\textbf{\textit{Our Theory:}} We extend Norman's concept but explicitly separate \emph{internal affordances} (what users perceive) from \emph{external affordances} (what actually exists in the environment). In our framework, perceived affordances correspond to internal affordances -- constructed within the internal environment using sensory inputs and prior experience. If a mismatch arises between the internal (perceived) and external (actual) affordance, feedback from interactions updates their internal models, gradually aligning perceptions with reality. This approach explains how users refine their interpretations over time, reducing errors in perceived affordances through learned experience.

\vspace{5px}
\noindent
\textbf{6. Misperceived Affordance}\\
\textbf{\textit{Definition:}} Misperceived affordance include \textit{false affordance} (when an object or interface appears to afford an action that is not possible) and \textit{hidden affordance} (when an action is possible but the user does not perceive it) \cite{10.1145/108844.108856}. \\
\textbf{\textit{Example:}} A decorative button that looks clickable but does nothing (false affordance) or a hidden shortcut that users fail to notice (hidden affordance). \\
\textbf{\textit{Our Theory:}} Misperceptions arise from a misalignment between internal and external affordances. Initially, users infer affordances from incomplete or misleading cues. Over time, however, through feedback and interaction, their internal models are refined, reducing the likelihood of false or hidden affordances.

\vspace{5px}
\noindent
\textbf{7. Signifiers}\\
\textbf{\textit{Definition:}} Signifiers are cues, such as symbols, words, or visual indicators, that help users understand what actions are possible or how to interact with an object or system \cite{norman2013design,norman1}. \\
%Unlike affordances, which are intrinsic to an object, signifiers are explicitly added to communicate how to use the object . \\
\textbf{\textit{Example:}} A ``push'' or ``pull'' sign on a door, or a visual indicator on a digital button showing it can be clicked. \\
\textbf{\textit{Our Theory:}} In our CR-based framework, signifiers are external cues that help users refine their internal models of affordances. Although the affordance itself is inferred through internal processes, signifiers simplify this process, reducing cognitive load by clarifying the intended action. In ambiguous situations, users can rely on signifiers to quickly infer affordances, improving the speed and accuracy of their decisions. 
%Over time, as users interact with the system, the need for signifiers decreases as their internal models become more robust and refined through feedback.

\vspace{5px}
\noindent
\textbf{8. Negative Affordance}\\
\textbf{\textit{Definition:}} Affordances that indicate the impossibility or undesirability of certain actions, acting as constraints that discourage specific behaviors \cite{masoudi2019review}. \\
%by lowering the confidence in success and the predicted utility of the action .\\
\textbf{\textit{Example:}} A locked door that clearly signals it cannot be opened, or a ``Danger: High Voltage'' sign that dissuades interaction due to potential harm.\\
\textbf{\textit{Our Theory:}} In our CR-based theory, negative affordances are perceived within the internal model as actions that either cannot be performed or carry low/negative predicted utility. Users identify cues, such as locks, barriers, or warning signs, which lower confidence in successfully performing the action and largely decrease the predicted utility. For instance, seeing a lock decreases confidence in opening the door, while encountering a ``Danger'' sign reduces utility due to the potential for harm. 
These negative affordances are learned and refined through reinforcement, where feedback from past interactions updates the internal model.

\vspace{5px}
\noindent
\textbf{9. Digital Affordance}\\
\textbf{\textit{Definition:}} Affordances in digital environments, where interactions are mediated through digital interfaces and design elements \cite{gaver1991technology, hartson2003cognitive}. \\
\textbf{\textit{Example:}} Clicking on an icon to open a file or swiping on a touchscreen to navigate. \\
\textbf{\textit{Our Theory:}} Digital affordances are detected through the user’s internal representation of the digital environment, which is constructed based on \textit{interface design cues} and prior experience with similar interactions. Users perceive digital affordances by relying on \textit{feature recognition} (e.g., visual cues indicating interactivity) and, when uncertain, simulate potential actions to predict the utility and confidence of the action. Over time, \textit{reinforcement learning} refines the user’s ability to recognize digital affordances, improving the efficiency of their interactions with virtual systems.

\vspace{5px}
\noindent
\textbf{10. Virtual Affordance}\\
\textbf{\textit{Definition:}} Affordances that arise within virtual environments rather than the physical world. Virtual affordances can be highly dynamic, flexible, and customized based on the design of the virtual environment \cite{jo2023affordance, bhargava2020revisiting}.\\
\textbf{\textit{Example:}} In a virtual reality game, players can teleport by pointing at a location, or picking up objects that appear solid but have no real-world counterpart.\\
\textbf{\textit{Our Theory:}} While there are no physical objects, virtual affordances can still be detected through constructing internal models of the virtual environment based on the sensory input (visual, haptic, audio) and feedback from the system. 
Although these affordances do not exist in the physical world, users can still learn and adapt to them through a learning process akin to the learning of digital affordances.
%Unlike digital affordances, which are tied to familiar interface elements like buttons and icons, virtual affordances rely more heavily on contextual and environmental cues unique to the virtual space. Users rely on sensory inputs, such as visual or haptic feedback from the virtual environment, and past experience to infer action possibilities. 
%Over time, through repeated interaction and feedback, users refine their internal representations, learning to navigate and manipulate virtual spaces with increasing efficiency.

\vspace{5px}
\noindent
\textbf{11. Sequential Affordance}\\
\textbf{\textit{Definition:}} Affordances that emerge through a sequence of actions, where one action leads to another \cite{gaver1991technology}. \\
\textbf{\textit{Example:}} Clicking through a series of prompts to complete a task (e.g., filling out a form and clicking ``submit''). \\
\textbf{\textit{Our Theory:}} Sequential affordances are perceived by simulating a series of \textit{hypothetical motion trajectories}. Users mentally simulate the results of each action in the sequence and predict the utility and confidence of each step. As users become more familiar with these sequential actions, their internal models improve, allowing them to navigate these sequences more efficiently, potentially even through feature matching. \textit{Reinforcement learning} strengthens their ability to recognize sequential affordances based on previous successes and failures.

\vspace{5px}
\noindent
\textbf{12. Learned Affordance}\\
\textbf{\textit{Definition:}} Affordances that are not immediately obvious but become apparent after learning or interactions \cite{cooper2001learning}. \\
\textbf{\textit{xample:}} Discovering that a long press on a touchscreen icon opens a submenu. \\
\textbf{\textit{Our Theory:}} Learned affordances are the result of repeated interactions with an object or environment. Through \textit{trial and error}, users gradually refine their understanding of the affordance. Feedback from both successful and failed attempts updates the user’s internal model, improving their ability to detect and use the affordance in future interactions. The learning process is governed by \textit{reinforcement learning}, with each interaction contributing to a more accurate internal representation.

\vspace{5px}

\noindent
\textbf{13. Affordance++}\\
\textbf{\textit{Definition:}} A concept that extends traditional affordances by integrating Electrical Muscle Stimulation (EMS), enabling objects to guide the user’s actions directly by stimulating their muscles to perform the appropriate motions \cite{lopes2015affordance}. \\
\textbf{\textit{Example:}} A spray can that uses EMS to make the user shake it correctly before spraying. \\
\textbf{\textit{Our Theory:}} Our CR-based theory complements Affordance++ by providing a cognitive framework for understanding how users refine their internal models through direct physical guidance. Initially, the user's internal model may simulate broader or incorrect hypothetical motion trajectories, but through repeated EMS-driven feedback, the correct motions are directly provided. 
Over time, users will rely less on hypothetical motion trajectories and instead directly associate specific cues with the correct actions, reinforced by the system-guided learning of these movements. For instance, after repeated interaction with an EMS-guided spray can, the user will automatically recognize the correct shaking motion without requiring further guidance from the system.

\vspace{5px}
\noindent
\subsection{Affordance in Social Contexts}
Finally, specific types of affordances have been introduced to explain the underlying mechanisms of behaviors in complex social dynamics. These affordances account for actions shaped by social norms, cultural expectations, and collective interactions. We demonstrate how CR-based theory can provide a unified foundation for explaining these social- or collective-based affordances.

\vspace{5px}

\noindent
\textbf{14. Social Affordance}\\
\textbf{\textit{Definition:}} Affordances that arise from interpersonal interactions or social dynamics, shaped by implicit social norms, behaviors, and expectations. These affordances often depend on interpreting social cues and understanding unspoken societal rules \cite{rietveld2013social}. \\
\textbf{\textit{Example:}} Extending a hand for a handshake as a greeting, or understanding not to tap someone’s forehead, as it could be interpreted as invasive or disrespectful. \\
\textbf{\textit{Our Theory:}} In our CR-based framework, social affordances are inferred by simulating social interactions within the internal environment. Users rely on prior experiences and social cues to infer potential outcomes in social contexts. For example, simulating a handshake generates positive predicted utility (e.g., maintaining social harmony), guiding the decision to act. Conversely, simulating an inappropriate gesture like tapping someone’s forehead leads to negative predicted utility (e.g., offending someone), guiding the decision not to act.
Over time, through feature recognition, the user learns to directly associate common terms or gestures, like a handshake, with friendly greetings and potentially offensive actions, such as tapping someone’s forehead, with negative social consequences. This allows for more automatic recognition of appropriate social actions without requiring deeper mental simulation. 
%Feedback from repeated social interactions refines the internal model, improving efficiency in navigating social dynamics and making predictions of utility more precise.

\vspace{5px}

\noindent
\textbf{15. Cultural Affordance} \\
\textbf{\textit{Definition:}} Affordances shaped by broader cultural norms, traditions, and societal values, influencing behavior based on learned expectations within a specific culture \cite{turner2005affordance}. \\ 
%Cultural affordances may differ significantly across societies and are often passed down through generations . \\
\textbf{\textit{Example:}} In Japan, bowing is an expected form of greeting, while in Western cultures, a handshake is more common. \\
\textbf{\textit{Our Theory:}} Cultural affordances are learned through interaction and feedback, where users internalize culturally specific actions and gestures. Initially, hypothetical motion trajectories allow exploration of possible actions, but over time, feature recognition helps users quickly associate actions with cultural meanings, reducing cognitive effort in familiar cultural situations.

\vspace{5px}

\noindent
\textbf{16. Collective Affordance}\\ 
\textbf{\textit{Definition:}} Affordances based on the combined capabilities of multiple individuals working together \cite{weichold2020collective, tison2021active}. \\
\textbf{\textit{Example:}} Lifting a heavy object that requires two or more people to coordinate their actions.\\ 
\textbf{\textit{Our Theory:}} With the \textit{internal environment} introduced in our CR-based theory, collective affordances are perceived when individuals not only create their own internal environment but also attempt to model the internal environments of others. This allows actors to infer each other's confidence and capabilities in performing a collective action. By leveraging \emph{hypothetical motion trajectories}, individuals can anticipate how others will contribute, enabling better predictions of the success of joint actions. 
Our theory provides a more comprehensive explanation of collective affordances, something traditional (direct-perception) affordance theories could not fully address.

\subsection{Summary}
Our \textit{CR-based affordance theory} provides a unified framework that explains how users perceive, learn, and refine their understanding of various types of affordances. Whether in physical, digital, social, or complex environments, our theory accounts for the role of \textit{internal representations}, \textit{simulated actions}, and \textit{learning through feedback}, providing a robust explanation of how affordances are detected and acted upon across a wide range of contexts. This unifying model enhances both the theoretical understanding of affordances and offers practical guidance for designers in developing more intuitive and responsive systems.

%% file: tex/7.discussion.tex
\section{Discussion}

Affordance is a key concept in HCI and design, explaining how users take actions and what defines intuitive interfaces. Yet, unclear definitions and varying interpretations have limited its practical use in design.
In this paper, we argue that to effectively explain affordance perception, unify existing theories, and understand the learning process, it is necessary to move beyond Gibson's direct-perception model. 
Our novel CR-based affordance theory posits that organisms infer affordances through internally constructed environments, shaped by bounded sensory inputs and prior experience. 
Affordance perception then becomes a problem of bounded optimality, where users evaluate the potential actions based on two key components: confidence (likelihood of success) and predicted utility (expected outcome value). 
Our theory introduces feature matching and hypothetical motion trajectories as core perception mechanisms, explaining how users simulate and refine affordances as a part of their decision-making. 
Unlike static views of affordance, we frame affordances as dynamic, continuously learned, and updated via reinforcement learning. 
Our CR-based theory not only bridges gaps between existing theories but also provides actionable insights for designing interfaces that better adapt to user feedback, enhancing both theoretical understanding and practical design applications.
We outline the key advantages offered by our theory and future works.

\subsection{Design Implications and Applications}
Our theory has significant implications for designers and practitioners in areas such as human-computer interaction (HCI), user experience (UX) design, and intelligent systems.

\paragraph{Improving Affordance Clarity} Our theory offers a more concrete framework for understanding how users perceive and learn affordances, helping designers create clearer, more intuitive interfaces. By unifying affordance perception across physical, digital, and virtual environments, our CR-based theory enables designers in different fields to apply consistent principles. This is particularly powerful in helping designers explain and communicate their design concepts based on the affordances perceived by users. Since our theory is tied to decision-making processes, it allows designers to predict and model how users choose between actions based on confidence and predicted utility, making affordance-driven design decisions more explicit and measurable.

\paragraph{Unified Framework} In the past, affordance theories have often been segmented by different types of interactions (e.g., sequential, digital, or social affordances), leading to fragmented terminology and conceptual divides in design. Our CR-based affordance theory bridges these gaps by providing a unified model for affordance perception and learning mechanisms across all categories. This unified framework simplifies communication and collaboration across multidisciplinary teams, as it offers a consistent, shared language to describe and understand user interactions. By reducing the complexity of affordance-related terms, the theory promotes clearer communication, enables more effective teamwork, and accelerates the design process.

\paragraph{Unraveling Natural Interactions} The concept of ``natural'' or ``intuitive'' interactions has long been a somewhat mysterious goal in design and HCI. We often recognize that some interfaces ``feel right'' without fully understanding why. Traditional theories, such as Gibson's, suggest that users instinctively perceive relevant affordances from the environment, but these explanations lack detail on the underlying process.
Our CR-based theory demystifies these interactions by proposing that what appears ``natural'' is actually the outcome of continuous learning and refinement. In this sense, what Gibson referred to as “direct perception” is actually the end result of a long learning process.
By recognizing this, designers can move beyond the vague goal of creating ``natural interactions'' and instead focus on crafting experiences that actively facilitate this learning process. Intuitive interfaces are those that provide clear feedback, align with users' past experiences, and help refine their internal models over time, resulting in interactions that feel effortless and instinctive.

\paragraph{Adaptive Systems}
Our CR-based affordance theory provides a foundation for designing future adaptive systems that respond to user interactions and experience levels. Adaptive systems can leverage the concept of affordance refinement over time, adjusting feedback or interface complexity based on a user's proficiency and familiarity with certain affordances. Novice users may receive additional cues or more explicit guidance, while expert users can engage with a more advanced version of the same system. This dynamic approach, which is grounded in users' evolving internal models, allows systems to cater to individual learning curves and optimize user experience through personalized adaptation. By providing the right level of affordance clarity and feedback at the right time, designers can create systems that not only facilitate learning but also enhance performance and user satisfaction as individuals become more skilled.

\paragraph{Human-AI Interaction}
Our CR-based theory highlights the importance of mutual affordance perception. Just as humans infer affordances when interacting with AI systems, future AI should be able to predict the affordances users will perceive, learn, and adopt. AI systems should assess whether introducing new features will enhance user experience or lead to confusion, tailoring interactions to user needs and learning curves. This mutual adaptation will foster more intuitive and personalized human-AI collaboration, making AI systems better at anticipating user preferences and abilities, and ultimately improving interaction quality.

\subsection{Limitation and Future Work}

\paragraph{Future Computational Affordance Models}

Traditional computational models view affordance perception as a visual classification task, using large datasets of labeled images to train deep neural networks with predefined motion labels \cite{nguyen2017object, AffordanceNet18, chuang2018learning}. 
However, these models fail to adapt to evolving capabilities or novel environments, limiting their real-world adaptability. Our CR-based affordance theory offers a more flexible approach. Instead of relying solely on static visual data, models based on this theory would incorporate (1) internal environment modeling from sensory inputs, (2) feature matching and hypothetical motion trajectories to infer affordances, and (3) continuous refinement through reinforcement learning.
To build such models, integrating Vision-Language Models (VLMs) \cite{radford2021learning, li2022blip} and model-based reinforcement learning (MBRL) \cite{chua2018deep, nagabandi2018neural} may be required. 
VLMs provide rich contextual descriptions, enhancing affordance interpretation, while MBRL ensures continuous adaptation by updating internal representations based on feedback. This allows robots to dynamically adapt their motions as capabilities evolve, which is essential in human-AI collaborations where robots must adjust actions based on shared environments. Predicting user interactions with affordances enables robots to offer more seamless assistance, improving collaboration.

\paragraph{Further Empirical Evaluations}

Our CR-based affordance theory provides a novel understanding of affordance perception, and we present a series of thought experiments to validate our theory. 
Future research should conduct further empirical testing, such as validating the theory across diverse environments with physical, digital, and virtual interfaces. These studies can continue unraveling detailed mechanisms of affordance learning, recognition, and corresponding decision-making processes. 
Additionally, simulations in robotic and AI systems would offer insights into how effectively computational models based on our theory can adapt to novel environments and explain human behaviors.

\paragraph{Extending CR-based Affordance with other Cognitive Architectures}

Future work should explore alternative cognitive architectures to further refine and extend the CR-based affordance theory. 
For instance, ACT-R's modular framework for modeling human cognition \cite{anderson2014atomic} could offer a more detailed understanding of how declarative and procedural knowledge contribute to affordance learning and decision-making, further providing more concrete underlying mechanisms of feature recognition and hypothetical motion trajectories. 
Furthermore, the Bayesian brain hypothesis \cite{knill2004bayesian} suggests that the brain is a probabilistic inference machine; This view could improve the understanding of affordance perception under uncertainty by dynamically updating confidence and utility based on sensory input and prior beliefs. 
Lastly, the Free Energy Principle \cite{friston2010free}, with its focus on minimizing prediction errors, aligns well with bounded rationality and could provide a robust foundation for modeling affordance inference, learning, and adaptation in highly dynamic and uncertain environments.

\paragraph{General Limitations of the Proposed Theory}

While the CR-based affordance theory provides a unifying framework for understanding affordance perception, it has several limitations that future research should address. 
First, our theory relies feature recognition and hypothetical motion trajectories to explain affordance perception, which may oversimplify the complexity of cognitive processes, especially in diverse and dynamic contexts. 
Furthermore, while the theory explains affordance adaptation through reinforcement learning, it does not fully capture the efficiency of human motion planning, which is often far faster and more adaptive than computational reinforcement learning \cite{botvinick2019reinforcement}. 
This discrepancy suggests more sophisticated and advanced learning strategies remain unexplored. 
Finally, the theory assumes that organisms can construct internal representations of the world, a capability that is challenging to computationally model and simulate. 
In particular, how humans construct such internal environments under uncertainty, especially in unfamiliar scenarios, requires further investigation.

\section{Conclusion}

This paper introduced a novel theory of affordance grounded in Computational Rationality, moving beyond the traditional direct-perception view to a model based on internal constructive processes. We argue that affordances are inferred within an internal environment, shaped by sensory inputs, experiences, and decision-making processes involving confidence and predicted utility.
Our theory offers a unified framework for various types of affordances, incorporating mechanisms including feature matching and hypothetical motion trajectories to explain how users dynamically perceive and refine affordances. We further emphasize that affordance perception is an evolving process, driven by reinforcement learning, rather than a static one.
The practical implications of our theory extend to HCI, UX design, and intelligent systems. By recognizing affordances as dynamic and learned, designers can create systems with a stronger theoretical foundation that better supports user learning, thereby enhancing performance across physical, digital, and virtual environments.
We anticipate our theory not only advances the conceptual understanding of affordances but also provides a foundation for future research and design practice in creating more adaptive, user-centered systems and interfaces.

%specifically, our theory offers improved affordance clarity for building future adaptive interfaces and natural interactions. 